\documentclass[12pt]{article}
\usepackage{amssymb,amsmath,epsfig}
\allowdisplaybreaks

\begin{document}

\title{\bf Stability of $d$-dimensional Gravastars with Variable Equation of State}
\author{M. Sharif \thanks {msharif.math@pu.edu.pk} and Faisal Javed
\thanks{faisaljaved.math@gmail.com}\\
Department of Mathematics, University of the Punjab,\\
Quaid-e-Azam Campus, Lahore-54590, Pakistan.}

\date{}
\maketitle

\begin{abstract}
In this paper, we are interested to explore stable configurations of
$d$-dimensional gravastars constructed from the interior
$d$-dimensional de Sitter and exterior $d$-dimensional
Schwarzschild/Reissner-Nordstr\"om (de Sitter) black holes through
cut and paste approach. We consider the linearized radial
perturbation preserving the original symmetries to explore their
stability by using three different types of matter distributions.
The resulting frameworks represent unstable structures for
barotropic, Chaplygin, and phantomlike models for every considered
choice of exterior geometries. However, matter contents with
variable equations of state have a remarkable role to maintain the
stability of gravastars. We conclude that stable structures of
gravastars are obtained only for generalized variable models with
exterior $d$-dimensional Schwarzschild/Reissner-Nordstr\"om-de
Sitter black holes.
\end{abstract}
\textbf{Keywords:} Gravastars; Black holes; Equation of state; Stability.\\
\textbf{PACS:} 04.20.-q; 04.70.Dy; 04.40.Dg; 04.50.Gh.

\section{Introduction}

A decade or more ago, Mazur and Mottola \cite{1,2} proposed a new
solution for the endpoint of a gravitational collapse. They
developed a cold compact object composed of an interior de Sitter
(dS) and exterior Schwarzschild geometry by applying the principle
of Bose-Einstein condensation to gravitational structures. These
geometries are partitioned by a phase boundary having a small and
finite thickness ($r_i-r_o=\delta$) which is also referred to as
thin-shell, where the inner and outer radii of gravastar are denoted
by $r_i$ and $r_o$, respectively. Consequently, the matter
distributions at these sections can be characterized through some
specific equations of state (EoS) given as
\begin{itemize}
\item Inner manifold: $0\leq r<r_i$, $p=-\sigma$,
\item Thin-shell: $r_i< r<r_o$, $p=\sigma$,
\item Outer manifold: $r_o<r$, $p=0=\sigma$.
\end{itemize}
The surface energy density and pressure of the matter contents are
represented by $\sigma$ and $p$, respectively. To achieve a stable
configuration, the concentration of matter at thin-shell plays a
vital role to overcome the rapidly expanding and collapsing nature
of thin-shell. The cut and paste technique \cite{2a} offers a
general formalism for thin-shell construction from the matching of
two distinct spacetimes at hypersurface. This formalism is used to
construct thin-shell gravastar from the matching of interior dS and
external Schwarzschild BH \cite{2}. This method is very useful in
avoiding the existence of central singularity and event horizon in
the developed geometrical structure. We have also studied thin-shell
stability as well as dynamics developed from cut and paste approach
\cite{2aa,2aaa}.

Gravastars have attracted many researchers to study their basic
framework using various methods. Visser and Wiltshire \cite{3}
discussed stable characteristics of gravastars by introducing
different EoS. Horvat and Ilijic \cite{3b} studied gravastar model
with exterior Schwarzschild and Schwarzschild-dS geometries. This
formalism is extended for inner dS and outer Reissner-Nordstr\" om
(RN) BH in \cite{4,5}. The physical characteristics like length,
energy contents, and entropy of (2+1)-dimensional charged free and
charged gravastars are investigated in \cite{6}. The charged
gravastar model with interior dS and exterior RN-dS spacetime is
constructed in \cite{6a}.

The study of the possible existence of bounded excursion gravastars
(in which shell radius oscillates between two radii) is an
interesting topic in the background of different exterior
geometries. The prototype gravastar model was introduced by Rocha
and his collaborators \cite{8} by using external Schwarzschild BH
and internal dS spacetime with stiff as well as dust fluid
distributions. They analyzed that the developed structure expresses
stable bounded excursion gravastars or collapsing behavior until BHs
are formed. Thus, in such complex models, the probability of the
existence of a gravastar model cannot be eliminated. This analysis
was also extended for inner dS and outer Schwarzschild-dS and RN BHs
in \cite{14,15}. It is found that the presence of a cosmological
constant in the exterior geometry leads to a stable configuration of
gravastar \cite{14}.

The stability of the gravastar model with a different type of matter
distribution through linearized radial perturbation is also an
interesting issue. Lobo and Arellano \cite{15cc} constructed
thin-shell gravastars with nonlinear electrodynamics. The stable
configuration of noncommutative thin-shell gravastar is investigated
by Lobo and Garattini \cite{10}. They noticed that stable regions
must exist near the predicted location of the event horizon.
\"{O}vg\"{u}n et al. \cite{11} investigated the stability of
gravastars constructed from the interior dS and exterior charged
noncommutative BHs. They concluded that the developed structure
shows stable behavior near the formation of the event horizon for
suitable values of the physical parameters. Recently, we have
constructed thin-shell gravastar from exterior regular as well as
charged Kiselve BHs and explored the stability of the developed
geometry through radial perturbation by considering variable EoS
\cite{12}.

A significant subject in general relativity and modified gravity
theories is the study of gravastars in higher dimensions. The class
of gravastar solutions as an alternative to $d$-dimensional BHs are
presented in \cite{14a}-\cite{14c}. A fascinating type of
traversable wormhole is the thin-shell kind, which is built by
grafting two equivalent manifolds together to form a geodesically
complete spacetime with a shell positioned at the junction
interface. Dias and Lemos \cite{13d} constructed $d$-dimensional
electrically charged thin-shell wormholes and observed their
stability. Eiroa and Simeone \cite{13g} studied the stability of
$d$-dimensional static shells through spherically symmetric radial
perturbation and found that stable regions are increased for higher
dimensions. The stable configurations of $d$-dimensional thin-shell
surrounded by quintessence through linearized radial perturbation
are analyzed in \cite{13gg} and its dynamics with massless and
massive scalar fields are observed in \cite{13j}.

In this paper, we study the stability of higher-dimensional
gravastars by taking as an alternative to $d$-dimensional
Schwarzschild, Schwarzschild-dS, RN and RN-dS BHs. The paper is
organized in the following format. Section \textbf{2} is devoted to
construct $d$-dimensional gravastars and consider Lanczos equations
to obtain matter contents of thin-shell. Section \textbf{3} is used
to discuss the stability of the developed structure filled with
barotropic and generalized variable EoS. We observe the stability of
gravastars for different $d$-dimensional BHs. In the last section,
we briefly discuss the outcomes.

\section{Geometry of $d$-Dimensional Gravastars}

We consider Israel thin-shell formalism for smooth matching of the
interior $d$-dimensional dS spacetime with different exterior
$d$-dimensional BH geometries at the hypersurface. We develop the
mathematical expression of the equation of motion of gravastar by
using Lanczos equations. This equation plays a vital role to express
the dynamical as well as stable configurations of the established
frameworks in the presence of different types of matter
distributions.

\subsection{Interior Geometry}

In general relativity, the dS geometry serves as the simplest
mathematical model which explains the physics of the current
scenario associated with the observed accelerated expansion of the
universe. It is a vacuum solution of the field equations with
negative pressure and positive vacuum energy density. The universe
itself behaves as asymptotically dS with cosmological confirmations.
The $d$-dimensional dS spacetime is a Lorentzian manifold having
maximum symmetrical configurations with positive scalar constant
curvature and is an analog of a $d$-sphere. The line element of the
$d$-dimensional dS spacetime can be written as \cite{12f}
\begin{eqnarray}\label{1}
ds^2_-&=&-\mathcal{H}(r_-)dt^2_-+\mathcal{H}(r_-)^{-1}
dr^2_-+r^2_-d\Omega^2_{d-2},
\end{eqnarray}
where the line element of $(d-2)$ unit sphere is denoted by
$d\Omega^2_{d-2}$ and its metric function is given as
\begin{eqnarray}\nonumber
\mathcal{H}(r_-)=1-\frac{r^2_-}{L_i^2},\quad \text{with}\quad
L_i=\sqrt{\frac{(d-1)(d-2)}{2\Lambda_i}},
\end{eqnarray}
here $\Lambda_i$ represents the cosmological constant and
cosmological event horizon at $r_{c-}=\alpha$. For different choices
of $\Lambda_i$, it leads to different spacetimes, i.e., dS for
$\Lambda_i>0$, anti dS (AdS) for $\Lambda_i<0$ and flat region for
$\Lambda_i=0$.

\subsection{Exterior Geometry}

To construct $d$-dimensional gravastar, we consider the exterior
spacetime as a $d$-dimensional BH. The line element of this
spacetime is \cite{12f}
\begin{eqnarray}\label{3}
ds^2_+&=&-\mathcal{F}(r_+)dt^2_++\mathcal{F}(r_+)^{-1}
dr^2_++r^2_+d\Omega^2_{d-2}.
\end{eqnarray}
Different choices of the metric function ($\mathcal{F}(r_+)$) leads
to different BHs.
\begin{itemize}
\item For $d$-dimensional Schwarzschild BH, it can be expressed as
\begin{eqnarray}\nonumber
\mathcal{F}(r_+)=1-\frac{2m}{r^{d-3}_+},
\end{eqnarray}
where $m$ is the mass of BH.
\item For $d$-dimensional Schwarzschild-dS BH, it yields
\begin{eqnarray}\nonumber
\mathcal{F}(r_+)=1-\frac{2m}{r^{d-3}_+}-\frac{r^2_+}{L_e^2},\quad
\text{with}\quad L_e=\sqrt{\frac{(d-1)(d-2)}{2\Lambda_e}},
\end{eqnarray}
where $\Lambda_e$ represents the cosmological constant of exterior
geometry.
\item For $d$-dimensional RN BH, it turns out to be
\begin{eqnarray}\nonumber
\mathcal{F}(r_+)=1-\frac{2m}{r^{d-3}_+}+\frac{Q^2}{r^{2(d-3)}_+},
\end{eqnarray}
where $Q$ is the charge distribution of the BH spacetime.
\item For $d$-dimensional RN-dS BH, it becomes
\begin{eqnarray}\nonumber
\mathcal{F}(r_+)=1-\frac{2m}{r^{d-3}_+}+\frac{Q^2}{r^{2(d-3)}_+}-\frac{r^2_+}{L_e^2}.
\end{eqnarray}
\end{itemize}

\subsection{Visser Cut and Paste Approach}

Visser's approach is the best mathematical technique to construct a
geometrical structure free from event horizons and singularities
through the joining of inner and outer spacetimes. We apply this
technique to obtain the geometrical structure of $d$-dimensional
gravastar. For this purpose, we match these spacetimes at the
timelike hypersurface ($\Sigma$) which is also known as
$(d-1)$-dimensional spacetime. The hypersurface is referred to as a
gravastar shell with radius $r=y(\tau)\equiv y$, where $\tau$ is the
proper time. To avoid the event horizons and singularities in the
developed structure, the shell radius $y$ must be greater than the
event horizons of the considered spacetimes. According to Israel
junction conditions, the line element of a hypersurface, inner and
outer manifolds are identical, i.e., $ds^2_-\cong ds^2_\Sigma \cong
ds^2_+$. The corresponding line elements (\ref{1}) and (\ref{3}) of
inner and outer manifolds at $r_+=r_-=y$ can be expressed as
\begin{eqnarray}\label{5}
ds^2_-&=&\left[-\mathcal{H}(y)+\mathcal{H}(y)^{-1}(dy/d\tau)^{2}
(d\tau/dt)^{2}\right]dt^2+y^2d\Omega^2_{d-2},\\\label{6a}
ds^2_+&=&\left[-\mathcal{F}(y)+\mathcal{F}(y)^{-1}(dy/d\tau)^{2}
(d\tau/dt)^{2}\right]dt^2+y^2d\Omega^2_{d-2}.
\end{eqnarray}
The line element of the hypersurface is given as
\begin{equation}\label{7}
ds^2_\Sigma=-d\tau^{2}+y^2d\Omega^2_{d-2},
\end{equation}
The proper time derivative of temporal coordinate is obtained by
comparing Eq.(\ref{7}) with Eqs.(\ref{5}) and (\ref{6a}). Hence, we
have
\begin{eqnarray}\label{8}
dt/d\tau=\dot{t}=\begin{cases}(\mathcal{H}(y)+
\dot{y}^2)^{1/2}/\mathcal{H}(y),\quad
\text{for the interior region}\\
(\mathcal{F}(y)+\dot{y}^2)^{1/2}/\mathcal{F}(y),\quad\text{for the
exterior region.}\end{cases}
\end{eqnarray}

The second Israel junction condition demonstrates discontinuity of
the extrinsic curvature at $\Sigma$, i.e.,
$[K^{a}_{b}]=K^{a}_{b(i)}-K^{a}_{b(e)}\neq0$. The matching of
considered spacetimes leads to the boundary surface if
$[K^{a}_{b}]=0$, otherwise it shows the existence of thin-shell. The
extrinsic curvature components can be defined as
\begin{equation}\label{9}
K_{ij}^{\pm}=-n_{\beta}^{\pm}\left(\frac{d^2y_{\pm}^\beta}
{d\xi^id\xi^j} +\Gamma^{\beta
\pm}_{\mu\nu}\frac{dy^{\mu}_{\pm}}{d\xi^{i}}\frac{dy^{\nu}_{\pm}}
{d\xi^{j}}\right),\quad i,j=0,2,3,\dotso,d,
\end{equation}
where $\xi^{i}$ denotes hypersurface coordinates and the outward
unit normals to the considered manifolds are given as
\begin{eqnarray}\nonumber
n_{\alpha}^{-}&=&\pm\mid g^{\mu\nu}\frac{\partial
\mathcal{H}(y)}{\partial y^\mu}\frac{\partial
\mathcal{H}(y)}{\partial y^\nu}\mid^{-\frac{1}{2}}\frac{\partial
\mathcal{H}(y)}{\partial y^\alpha},\quad
\alpha=0,1,2,\dotso,d,\\\nonumber n_{\alpha}^{+}&=&\pm\mid
g^{\mu\nu}\frac{\partial \mathcal{F}(y)}{\partial
y^\mu}\frac{\partial \mathcal{F}(y)}{\partial
y^\nu}\mid^{-\frac{1}{2}}\frac{\partial \mathcal{F}(y)}{\partial
y^\alpha}.
\end{eqnarray}
The corresponding $(t,t)$ and $(\theta,\theta)$ components of
extrinsic curvature become
\begin{eqnarray}\label{10}
K^{\tau}_\tau&=&\begin{cases}(\mathcal{H}'(y)+2\ddot{y}^2)/
(\mathcal{H}(y)
+\dot{y}^2)^{1/2},\quad \text{for interior region}\\
(\mathcal{H}'(y)+2\ddot{y}^2)/(\mathcal{F}(y)
+\dot{y}^2)^{1/2},\quad \text{for exterior region}\end{cases}
\\\label{11} K^{\theta}_\theta&=&\begin{cases}(\mathcal{H}(y)
+\dot{y}^2)^{1/2}/y,\quad \text{for interior region}\\
(\mathcal{F}(y) +\dot{y}^2)^{1/2}/y,\quad \text{for exterior
region}\end{cases}
\end{eqnarray}
where dash represents derivative to the shell radius $y$.

The discontinuity of extrinsic curvature is due to the presence of a
thin layer of matter surface at $\Sigma$. Such type of matter
distribution greatly affects the dynamics and stability of the
established frameworks. The stress-energy tensor components of
matter contents can be evaluated through the reduced form of the
field equations at $\Sigma$. These equations are known as Lanczos
equations and are given as
\begin{equation}\label{12}
8\pi S_{i}^j=\delta_{i}^j K-[K_{i}^j],
\end{equation}
where $S_{i}^j$ represents the stress-energy tensor and
$\delta_{i}^j$ is known as the Kronecker delta. The stress-energy
tensor for perfect fluid has the form
\begin{equation}\label{13}
S_{i}^j=v_iv^j\left(\rho+p\right)+\delta^j_ip,
\end{equation}
where $v_i$, $\rho$ and $p$ represent the shell velocity, energy
density and surface pressure of thin layer of matter at $\Sigma$,
respectively. For the considered manifolds, the Lanczos equations
can be expressed as
\begin{eqnarray}\label{14} \rho(y)&=&-\frac{d-2}{8\pi y}
\left\{\sqrt{\dot{y}^2+\mathcal{F}(y)}-\sqrt{\dot{y}^2+
\mathcal{H}(y)}\right\},
\\\nonumber p(y)&=&\frac{1}{8\pi y}\left\{\frac{y\mathcal{F}'(y)+2
y \ddot{y}+2 (d-3) \left(\dot{y}^2+\mathcal{F}(y)\right)}
{\sqrt{\dot{y}^2+
\mathcal{F}(y)}}\right.\\\label{15}&-&\left.\frac{y
\mathcal{H}'(y)+2 y \ddot{y}+2 (d-3)
\left(\dot{y}^2+\mathcal{H}(y)\right)}
{\sqrt{\dot{y}^2+\mathcal{H}(y)}} \right\}.
\end{eqnarray}
It is noted that the shell does not move at equilibrium shell radius
$y=y_0$ in the radial direction. Hence, the proper time derivative
of shell radius at $y=y_0$ vanishes, i.e., $\dot{y}_0=0=\ddot{y}_0$.
The corresponding expressions of energy density and surface pressure
at $y=y_0$, yield
\begin{eqnarray}\label{16}
\rho(y_0)&=&-\frac{d-2}{8\pi y_0}
\left\{\sqrt{\mathcal{F}(y_0)}-\sqrt{\mathcal{H}(y_0)}\right\},
\\\label{17} p(y_0)&=&\frac{y_0 \mathcal{F}'(y_0)+2 (d-3)\mathcal{F}(y_0)}
{8\pi y_0\sqrt{\mathcal{F}(y_0)}}-\frac{y_0 \mathcal{H}'(y_0)+2
(d-3)\mathcal{H}(y_0)} {8\pi y_0\sqrt{\mathcal{H}(y_0)}} .
\end{eqnarray}
The equation of motion of the shell can be obtained from
Eq.(\ref{14}) as
\begin{eqnarray}\label{18}
\frac{1}{2}\dot{y}^2 + \Pi(y)=0,
\end{eqnarray}
where
\begin{eqnarray}\label{19}
\Pi(y)&=&-\frac{8 \pi ^2 y^2 \sigma ^2}{(d-2)^2}-\frac{(d-2)^2
\mathcal{Z}^2(y)}{512 \pi ^2 y^2 \sigma
^2}+\frac{\mathcal{W}(y)}{4},
\end{eqnarray}
and the functions $\mathcal{Z}(y)$ and $\mathcal{W}(y)$ are defined
as
\begin{eqnarray}\nonumber
\mathcal{Z}(y)=\mathcal{H}(y)-\mathcal{F}(y),\quad
\mathcal{W}(y)=\mathcal{H}(y)+\mathcal{F}(y).
\end{eqnarray}

\section{Stability Analysis}

In this section, we discuss the stability of the developed structure
by using linearized radial perturbation. We consider three different
choices of matter distributions, i.e., barotropic EoS, generalized
Chaplygin variable (GCV) and generalized phantomlike variable (GPV)
models. The last two models depending on the shell radius of the
developed structure are also known as variable EoS. We develop
respective expressions of the effective potential of gravastar for
these EoS. We then explore the behavior of effective potential to
observe the final fate of the developed structure. It is mentioned
here that the behavior of gravastar geometry (stable and unstable
configurations) can be evaluated through the graphical behavior of
effective potential. If the potential function and its first
derivative vanish at $y=y_0$, then it shows stable behavior for
$\Pi''(y_0)>0$, unstable structure if $\Pi''(y_{0})<0$ and
unpredictable if $\Pi''(y_{0})=0$ \cite{20a}. We use linear radial
perturbation of the effective potential about equilibrium shell
radius up to second-order terms to explore the stable
configurations.

The Taylor series of $\Pi(y)$ about $y=y_0$ is given as
\begin{equation}\nonumber
\Pi(y)=\Pi(y_{0})+\Pi'(y_{0})(y-y_{0})+\frac{1}{2}
\Pi''(y_{0})(y-y_{0})^2+O[(y-y_{0})^3].
\end{equation}
It is found that $\Pi(y_0)=0=\Pi'(y_0)$, hence the above equation
becomes
\begin{equation}\label{20}
\Pi(y)=\frac{1}{2}(y-y_{0})^2\Pi''(y_{0}).
\end{equation}
The relation between surface stresses of thin-shell can be
determined through the conservation equation as
\begin{eqnarray}\label{21}
4\pi\frac{d}{d\tau}( y^2 \sigma)+4\pi p(\sigma,y)\frac{d
y^2}{d\tau}=0,
\end{eqnarray}
which can be written as
\begin{eqnarray}\label{22}
\frac{d\sigma}{dy}=-\frac{2}{y}(\sigma+p(\sigma,y)).
\end{eqnarray}
The derivative of the above equation with respect to the shell
radius becomes
\begin{equation}\label{23}
\frac{d^2\sigma}{dy^2}=\frac{p(\sigma,y)+\sigma}{y^2}
\left(2+\vartheta^2\right),
\end{equation}
where the EoS parameter can be expressed as
$\vartheta^2=dp/d\sigma$.

To investigate stability, we choose a specific expression of
$p(\sigma,y)$ which leads to different types of matter
distributions. We begin with linear EoS, $p=\eta \sigma$, where
$\eta$ is the barotropic EoS parameter. This gives a linear
relationship between surface energy density and pressure of matter
contents located at thin-shell. The corresponding solution of the
conservation equation (\ref{22}) yields
\begin{equation}\label{24}
\sigma=\left(y_{0}/y\right)^{2(1+\eta)}\sigma_{0}.
\end{equation}
The respective expression of $\Pi(y)$ turns out to be
\begin{equation}\label{25}
\Pi(y)=\frac{1}{4}\mathcal{W}(y)-\frac{(d-2)^2 \mathcal{Z}(y)^2}{512
\pi ^2 y^2 \sigma_0^2} \left(\frac{y_0}{y}\right)^{-4
(\eta+1)}-\frac{8 \pi ^2 y^2
\sigma_0^2}{(d-2)^2}\left(\frac{y_0}{y}\right)^{4 (\eta+1)}.
\end{equation}
It is found that $\Pi(y_0)=0$ and $\Pi'(y_0)$ can be written as
\begin{eqnarray}\nonumber
\Pi'(y_0)&=&\frac{16 \pi ^2y_0\sigma_{0}^2}{(d-2)^2} (2 \eta+1)
-\frac{(d-2)^2 \mathcal{Z}(y_0)}{256 \pi ^2 \sigma_{0}^2 y_0^3}
\left(2 \eta \mathcal{Z}(y_0)+y_0
\mathcal{Z}'(y_0)+\mathcal{Z}(y_0)\right)\\\nonumber&+&
\frac{\mathcal{W}'(y_0)}{4},
\end{eqnarray}
which vanishes only if
\begin{equation}\label{26}
\eta=\frac{\frac{(d-2)^2 \mathcal{Z}(y_0)^2}{256 \pi ^2 \sigma_{0}^2
y_0^3}+\frac{(d-2)^2 \mathcal{Z}(y_0) \mathcal{Z}'(y_0)}{256 \pi ^2
\sigma_{0}^2 y_0^2}-\frac{16 \pi ^2 \sigma_{0}^2
y_0}{(d-2)^2}-\frac{\mathcal{W}'(y_0)}{4}}{\frac{32 \pi ^2
\sigma_{0}^2 y_0}{(d-2)^2}-\frac{(d-2)^2 \mathcal{Z}(y_0)^2}{128 \pi
^2 \sigma_{0}^2 y_0^3}}.
\end{equation}
Hence, we have
\begin{eqnarray}\nonumber
\Pi''(y_0)&=&-\left(-64 \pi^2 \sigma_{0}^2 y_0^4 \left((d-2)^2
\mathcal{W}''(y_0)-64 \pi ^2 \sigma_{0}^2 (2 \eta+1) (4
\eta+3)\right)\right.\\\nonumber&+&\left.(d-2)^4 \left(8 \eta^2+6
\eta+1\right) \mathcal{Z}(y_0)^2+(d-2)^4 y_0 \mathcal{Z}(y_0)
\left(y_0 \mathcal{Z}''(y_0)\right.\right.
\\\nonumber&+&\left.\left.(8 \eta+4)
\mathcal{Z}'(y_0)\right)+(d-2)^4 y_0^2
\mathcal{Z}'(y_0)^2\right)\\\label{27}&\times&(256 \pi ^2 (d-2)^2
\sigma_{0}^2 y_0^4)^{-1}.
\end{eqnarray}
This equation is used to explore the stable/unstable geometrical
configurations of thin-shell gravastars filled with the barotropic
type fluid distributions.

We also obtain dynamical equations of the developed configurations
filled with fluid distribution for GCV EoS given as
$p=\frac{1}{y^n}\frac{\beta}{\sigma}$, where $\beta$ and $n$ are the
real constants \cite{20,20a}, for $n=0$, the Chaplygin gas model is
recovered \cite{21}. The corresponding solution is
\begin{equation}\label{28}
\sigma^2=\frac{(n-4)\sigma_{0}^2y_{0}^{n+4}y^n+4\beta
y^4y_{0}^n-4\beta y^ny_{0}^4}{y^{n+4}y_{0}^{n}(n-4)}.
\end{equation}
Consequently, the effective potential leads to
\begin{eqnarray}\nonumber
\Pi(y)&=&\frac{1}{4}\mathcal{W}(y)-\frac{(d-2)^2}{512 \pi ^2}\frac{
(n-4) y^{n+2} \mathcal{Z}(y)^2 y_0^n}{\left(y_0^4 y^n \left((n-4)
\sigma_{0}^2 y_0^n-4 \beta\right)+4 y^4 \beta
y_0^n\right)}\\\label{29}&-&\frac{8 \pi ^2 y^{-n-2}
y_0^{-n}}{(d-2)^2 (n-4)} \left(y_0^4 y^n \left((n-4) \sigma_{0}^2
y_0^n-4 \beta\right)+4 y^4 \beta y_0^n\right).
\end{eqnarray}
At equilibrium shell radius, the effective potential vanishes and
its first derivative with respect to shell radius is given as
\begin{eqnarray}\nonumber
\Pi'(y_0)&=&\left(-2 (d-2)^4 \beta \mathcal{Z}(y_0)^2-(d-2)^2
\sigma_{0}^2 y_0^n \left((d-2)^2 \mathcal{Z}(y_0) \left(y_0
\mathcal{Z}'(y_0)\right.\right.\right.\\\nonumber&+&\left.\left.
\left.\mathcal{Z}(y_0)\right)-64 \pi ^2 \sigma_{0}^2 y_0^3
\mathcal{W}'(y_0)\right)+4096 \pi ^4 \sigma_{0}^4 y_0^4 \left(2
\beta+\sigma_{0}^2 y_0^n\right)\right)\\\nonumber&\times&(256
y^{n+3}_0 \pi ^2 (d-2)^2 \sigma_0^4)^{-1}.
\end{eqnarray}
For $\Omega'(y_0)=0$, we obtain
\begin{eqnarray}\nonumber
\beta&=&\frac{-\sigma_{0}^2 y_0^n}{8192 \pi ^4 \sigma_{0}^4 y_0^4-2
(d-2)^4 \mathcal{Z}(y_0)^2}\left(64 \pi ^2 \sigma_{0}^2 y_0^3
\left((d-2)^2
\mathcal{W}'(y_0)\right.\right.\\\label{6}&+&\left.\left.64 \pi ^2
\sigma_{0}^2 y_0\right)+(d-2)^4 (-y_0) \mathcal{Z}(y_0)
\mathcal{Z}'(y_0)-(d-2)^4 \mathcal{Z}(y_0)^2\right),
\end{eqnarray}
and
\begin{eqnarray}\nonumber
\Pi''(y_0)&=&\frac{y_0^{-2 (n+2)}}{256 \pi ^2 (d-2)^2
\sigma_{0}^6}\left(-(d-2)^4 \mathcal{Z}(y_0)^2 \left(16 \beta^2-2
\beta (n-7) \sigma_{0}^2
y_0^n\right.\right.\\\nonumber&+&\left.\left.\sigma_{0}^4 y_0^{2
n}\right)-(d-2)^4 \sigma_{0}^2 y_0^{n+1} \mathcal{Z}(y_0) \left(8
\sigma_{0} \mathcal{Z}'(y_0)+\sigma_{0}^2 y_0^n \left(y_0
\mathcal{Z}''(y_0)\right.\right.\right.\\\nonumber&+&
\left.\left.\left.4 \mathcal{Z}'(y_0)\right)\right)+64 \pi ^2
\sigma_{0}^6 y_0^{n+4} \left((d-2)^2 y_0^n \mathcal{W}''(y_0)-64 \pi
^2 \left(2 \beta
(n+1)\right.\right.\right.\\\label{30}&+&\left.\left.\left.3
\sigma_{0}^2 y_0^n\right)\right)-(d-2)^4 \sigma_{0}^4 y_0^{2 n+2}
\mathcal{Z}'(y_0)^2\right).
\end{eqnarray}
This is used to analyze stability of thin-shell gravastars filled
with matter which follows GCV EoS.

Finally, we take GPV EoS, i.e., $p=\frac{\lambda\sigma}{y^n}$, where
$\lambda$ and $n$ are real constants \cite{20,20a}. For $n=0$,
phantomlike EoS is obtained \cite{22}. The respective solution of
conservation equation gives
\begin{equation}\label{31}
\sigma=y_{0}^2y^{-2}\sigma_{0}e^{\frac{
\lambda\left(y_0^{-n}-y^{-n}\right)}{n}},
\end{equation}
and it follows that
\begin{equation}\label{32}
\Pi(y)=\frac{1}{4}\mathcal{W}(y)-\frac{y^2 (d-2)^2
\mathcal{Z}(y)^2}{512 \pi ^2 \sigma_{0} y_0^4} e^{\frac{2 \lambda
\left(y_0^{-n}-y^{-n}\right)}{n}}-\frac{8 \pi ^2
\sigma_{0}}{y_0^{-4}y^2}\frac{ e^{\frac{2\lambda
\left(y^{-n}-y_0^{-n}\right)}{n}}}{ (d-2)^2}.
\end{equation}
Also, we have
\begin{eqnarray}\nonumber
\Pi'(y_0)&=&\frac{y_0^{-n-3}(d-2)^{-2} }{256 \pi ^2
\sigma_{0}}\left(64 \pi ^2 \sigma_0 y_0^3 \left(64 \pi ^2 \sigma_{0}
y_0 \left(\lambda+y_0^n\right)+(d-2)^2 y_0^n
\mathcal{W}'(y_0)\right.\right.\\\nonumber&-&\left.\left.(d-2)^4
y_0^{n+1} \mathcal{Z}(y_0) \mathcal{Z}'(y_0) \right)-(d-2)^4
\mathcal{Z}(y_0)^2 \left(\lambda+y_0^n\right) \right).
\end{eqnarray}
It is found that $\Pi(y_0)=0$ and by taking $\Pi'(y_0)=0$, we obtain
\begin{eqnarray}\nonumber
\lambda&=&y_0^n \left(-64 \pi ^2 \sigma_0 y_0^3 \left((d-2)^2
\mathcal{W}'(y_0)+64 \pi ^2 \sigma_0 y_0\right)+(d-2)^4 y_0
\mathcal{Z}(y_0) \mathcal{Z}'(y_0)\right.\\\nonumber&+&\left.(d-2)^4
\mathcal{Z}(y_0)^2\right)\left(4096 \pi ^4 \sigma_0^2 y_0^4-(d-2)^4
\mathcal{Z}(y_0)^2\right)^{-1},
\end{eqnarray}
and hence
\begin{eqnarray}\nonumber
\Pi''(y_0)&=&\frac{y_0^{-2 (n+2)} \pi ^{-2}}{256 (d-2)^2
\sigma_0}\left(-(d-2)^4 y_0^{2 n+2} \mathcal{Z}'(y_0)^2-(d-2)^4
\mathcal{Z}(y_0)^2 \left(2
\lambda^2\right.\right.\\\nonumber&-&\left.\left.\lambda (n-3)
y_0^n+y_0^{2 n}\right) +64 \pi ^2 \sigma_0 y_0^4 \left((d-2)^2
y_0^{2 n}
\mathcal{W}''(y_0)\right.\right.\\\nonumber&-&\left.\left.64 \pi ^2
\sigma_0 \left(2 \lambda^2+\lambda (n+5) y_0^n+3 y_0^{2
n}\right)\right)-(d-2)^4 y_0^{n+1} \mathcal{Z}(y_0) \left(4
\left(\lambda\right.\right.\right.\\\label{33}&+&\left.\left.\left.y_0^n\right)
\mathcal{Z}'(y_0)+y_0^{n+1} \mathcal{Z}''(y_0)\right)\right).
\end{eqnarray}

In the following, we study the influence of physical parameters on
the stability of the structure through graphs.

\subsection{Different Choices of Exterior Geometries}

We are interested to explore the stability of the developed
structure in the background of different choices of exterior BH
spacetimes, i.e., $d$-dimensional Schwarzschild (dS) and RN (dS)
spacetimes. The possible existence of stable bounded excursion
gravastars for these exterior geometries have been explored in
four-dimensional spacetimes \cite{8}-\cite{15}. We develop the
respective equations of motion of thin-shell filled with barotropic,
GCV, and GPV EoS. We use a graphical approach to discuss the second
derivative of the effective potential for suitable values of
physical parameters. The stable characteristics of gravastar for
different values of $d$ are shown in Figures \textbf{1-12}. Here, we
use $d=4$ (blue), $d=6$ (red), $d=8$ (brown), $d=10$ (green).

\subsubsection{$d$-dimensional Schwarzschild BH}

In the background of $d$-dimensional Schwarzschild BH, we have
\begin{eqnarray}\nonumber
\mathcal{Z}(y)&=&2 m y^{3-d}-\frac{2 y^2 \Lambda _i}{(d-2)
(d-1)},\\\nonumber \mathcal{W}(y)&=&-\frac{2 y^2 \Lambda _i}{(d-2)
(d-1)}-2 m y^{3-d}+2.
\end{eqnarray}
Using the above expressions in Eqs.(\ref{27}), (\ref{30}) and
(\ref{33}), we obtain $\Pi''(y_0)$ for $d$-dimensional Schwarzschild
BH and thin-shell filled with matter satisfying barotropic, GCV and
GPV EoS, respectively. The developed structure represents the
unstable configuration for every choice of fluid distributions as
shown in Figures \textbf{1-3}. This shows that the choice of
interior $d$-dimensional dS and exterior $d$-dimensional
Schwarzschild BH leads to an unstable structure of gravastar.
\begin{figure}\centering
\epsfig{file=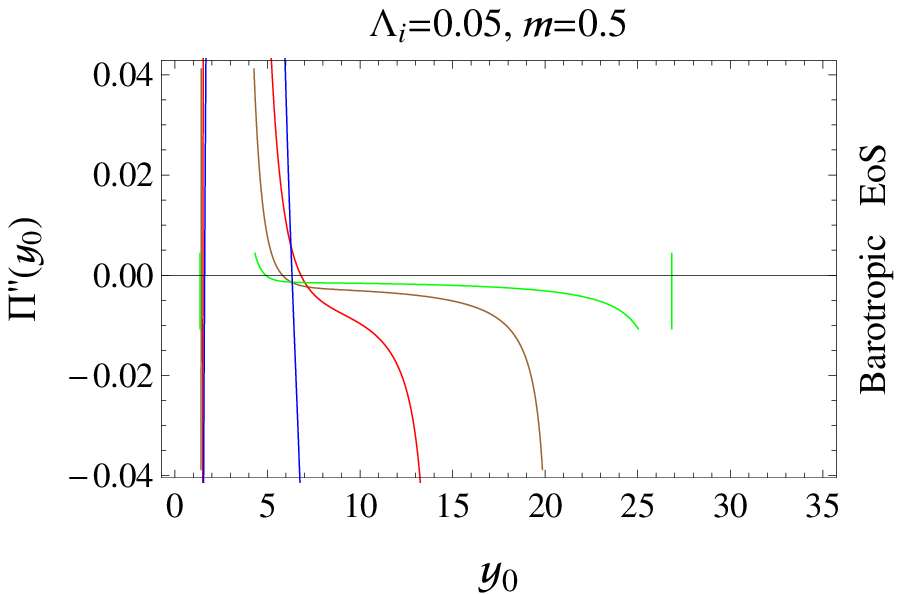,width=.5\linewidth} \caption{Unstable
configuration of $d$-dimensional gravastar for barotropic EoS with
exterior $d$-dimensional Schwarzschild BH. The blue, red, brown and
green colors represent the behavior of $\Pi''(y_0)$ for
$d=4,6,8,10$, respectively. It is noted that the developed structure
shows unstable configuration as $\Pi''(y_0)<0$ for every choice of
physical parameter.}
\epsfig{file=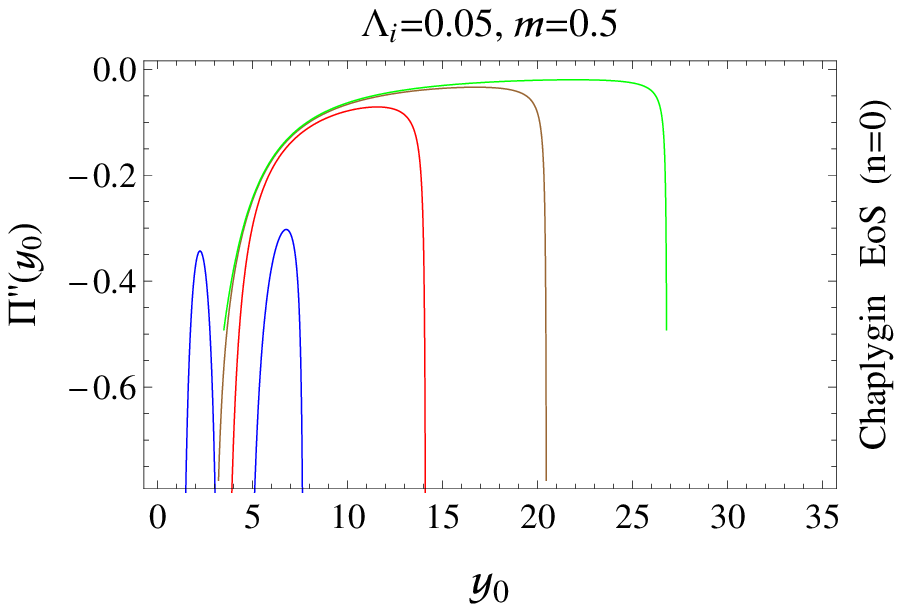,width=.5\linewidth}\epsfig{file=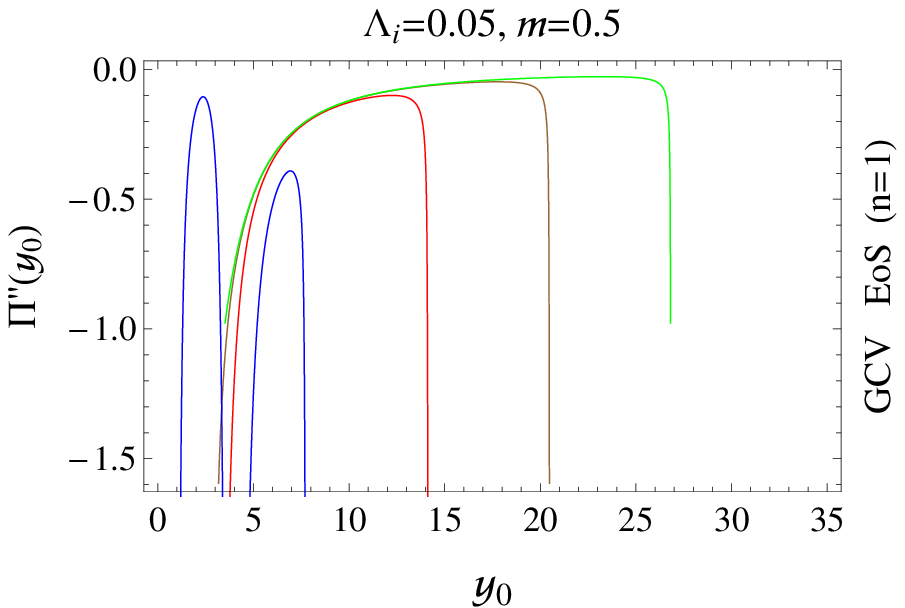,width=.5\linewidth}
\caption{Unstable configuration of $d$-dimensional gravastar for
Chaplygin and GCV EoS with exterior $d$-dimensional Schwarzschild
BH.}
\epsfig{file=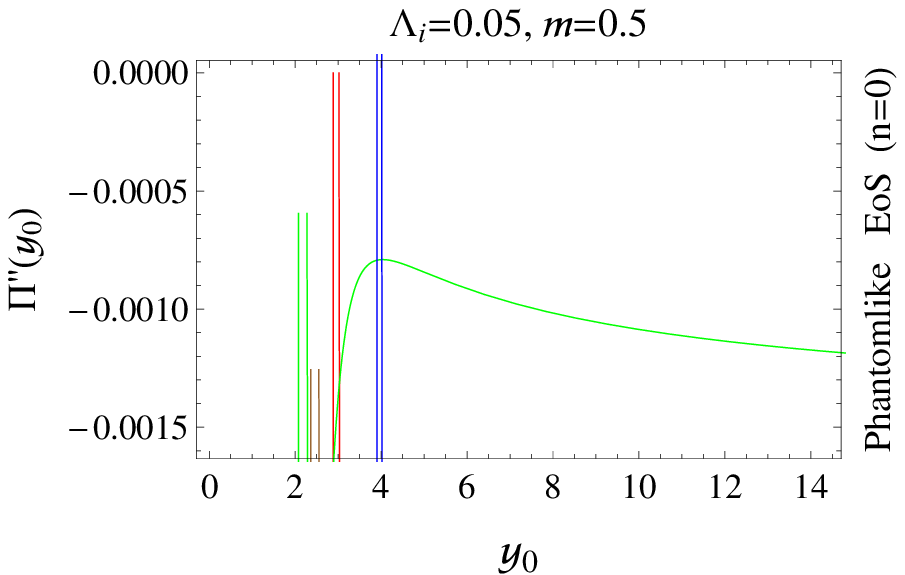,width=.5\linewidth}\epsfig{file=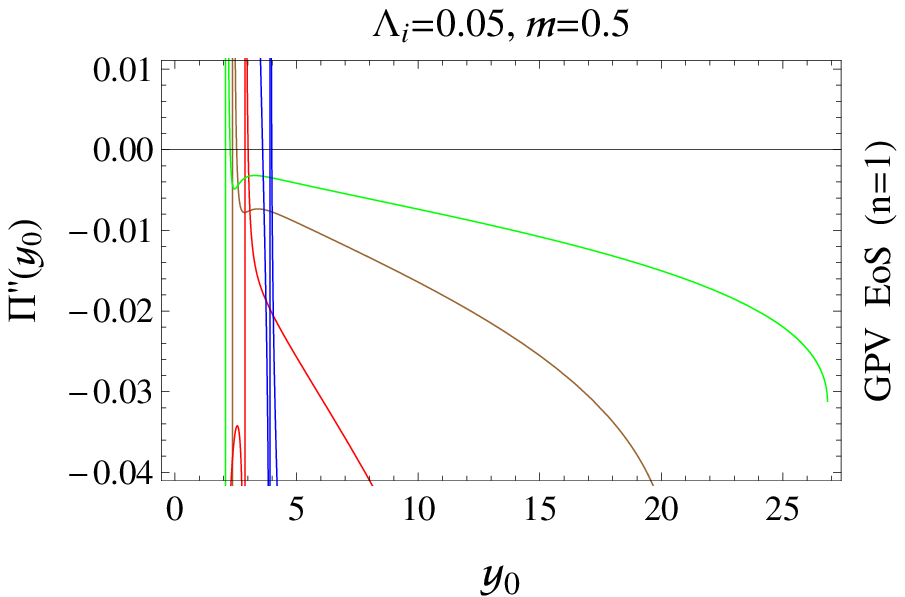,width=.5\linewidth}
\caption{Unstable configuration of $d$-dimensional gravastar for
phantomlike and GPV EoS with exterior $d$-dimensional Schwarzschild
BH.}
\end{figure}

\subsubsection{$d$-dimensional Schwarzschild-dS BH}

For $d$-dimensional Schwarzschild-dS BH, it follows that
\begin{eqnarray}\nonumber
\mathcal{Z}(y)&=&\frac{2 y^2 \Lambda _e}{(d-2) (d-1)}-\frac{2 y^2
\Lambda _i}{(d-2) (d-1)}+2 m y^{3-d},\\\nonumber
\mathcal{W}(y)&=&-\frac{2 y^2 \Lambda _e}{(d-2) (d-1)}-\frac{2 y^2
\Lambda _i}{(d-2) (d-1)}-2 m y^{3-d}+2.
\end{eqnarray}
Similarly, we obtain the respective expressions of $\Pi''(y_0)$ for
$d$-dimensional Schwarzschild-dS gravastar with different types of
matter contents by using the above functions in Eqs.(\ref{27}),
(\ref{30}) and (\ref{33}). Here, the gravastar structure shows
unstable behavior initially then represents stable structure for
barotropic type fluid distribution (Figure \textbf{4}). For
Chaplygin and phantomlike EoS, it represents unstable behavior (left
plots of Figures \textbf{5} and \textbf{6}) while GCV and GPV EoS
lead to stable configuration of the developed structure (right plots
of Figures \textbf{5} and \textbf{6}). Hence, the exterior
$d$-dimensional Schwarzschild-dS BH is suitable for the stable
geometry of gravastars than $d$-dimensional Schwarzschild BH.
\begin{figure}\centering
\epsfig{file=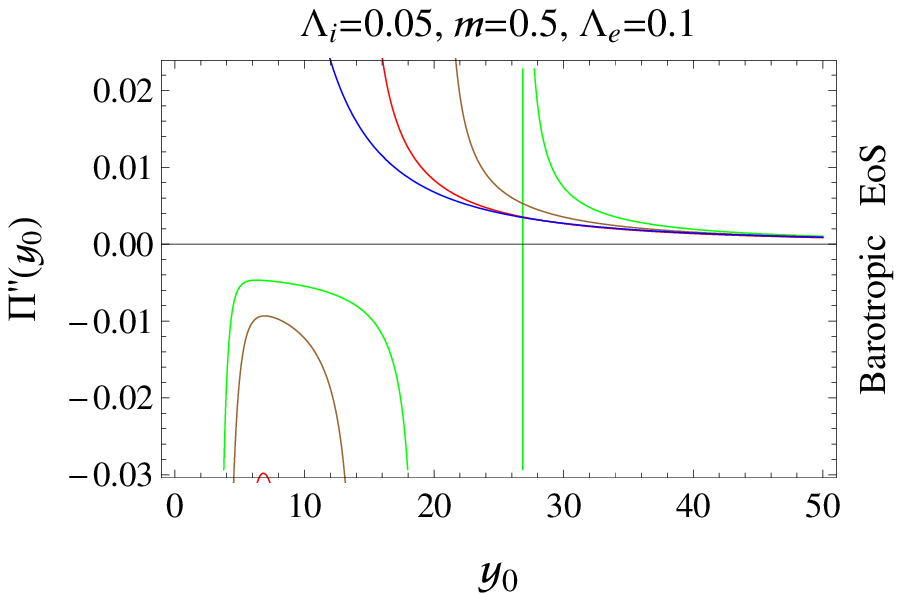,width=.5\linewidth} \caption{Stability of
$d$-dimensional gravastar for barotropic EoS with exterior
$d$-dimensional Schwarzschild-dS BH. It represents the initially
unstable and then stable configuration of the developed structure.}
\epsfig{file=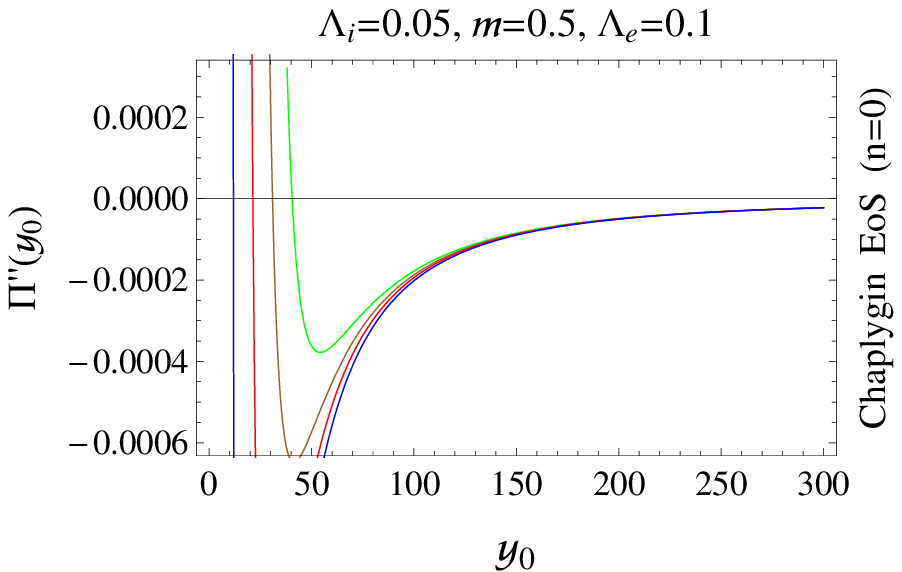,width=.5\linewidth}\epsfig{file=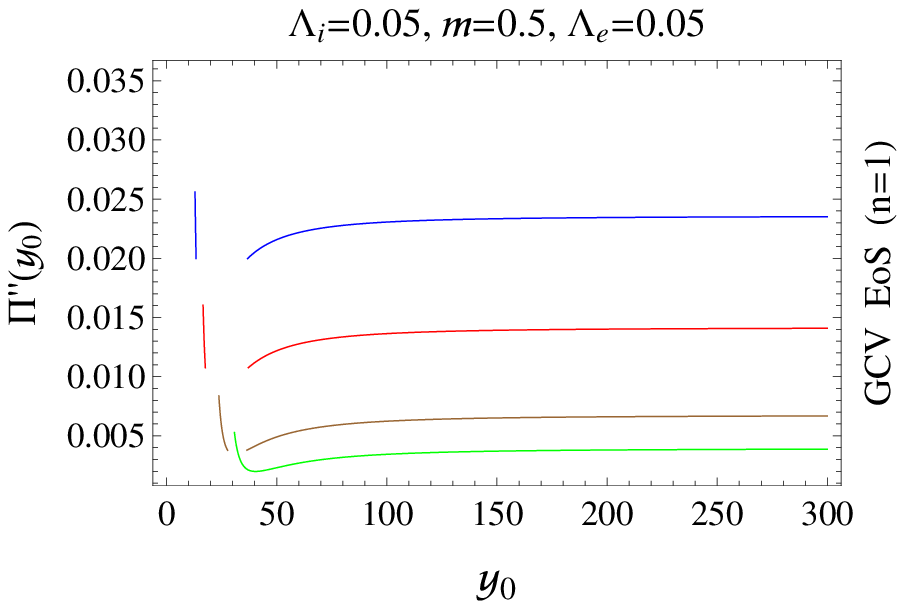,width=.5\linewidth}
\caption{Stability of $d$-dimensional gravastar for Chaplygin and
GCV EoS with exterior $d$-dimensional Schwarzschild-dS BH. Left plot
shows unstable structure for Chaplygin model and right plot
represents stable geometry for GCV EoS.}
\epsfig{file=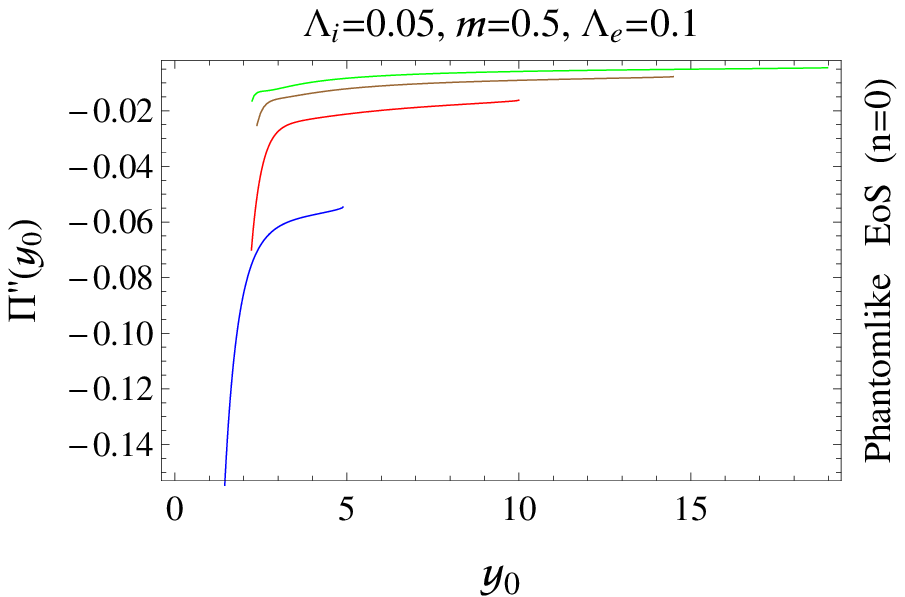,width=.5\linewidth}\epsfig{file=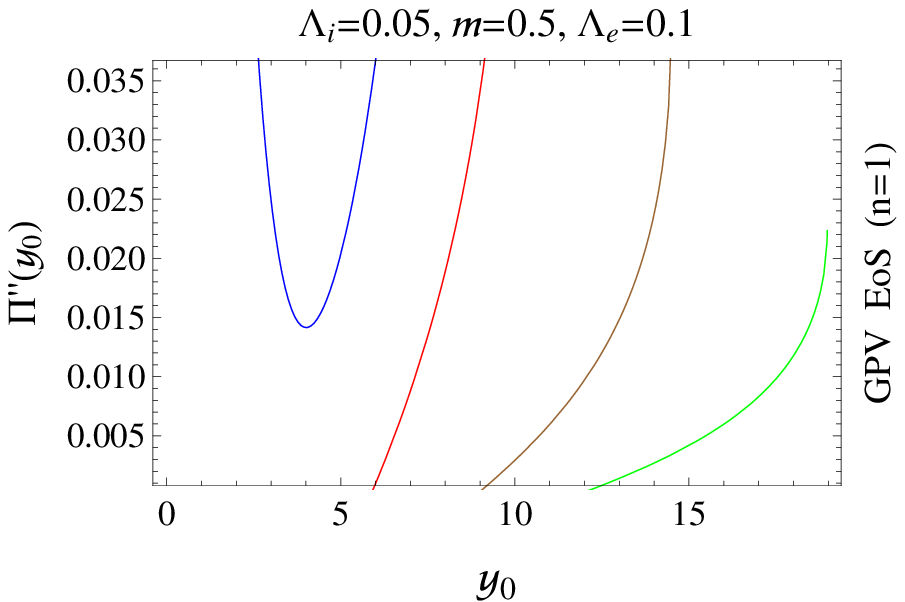,width=.5\linewidth}
\caption{Stability of $d$-dimensional gravastar for phantomlike and
GPV EoS with exterior $d$-dimensional Schwarzschild-dS BH. Left plot
indicates unstable configuration for phantomlike EoS while right
plot expresses stable structure for GPV EoS.}
\end{figure}

\subsubsection{$d$-dimensional Reissner-Nordstr\"om BH}

For $d$-dimensional RN BH, we have
\begin{eqnarray}\nonumber
\mathcal{Z}(y)&=&-\frac{2 y^2 \Lambda _i}{(d-2) (d-1)}+2 m
y^{3-d}-Q^2 y^{-2 (d-3)},\\\nonumber \mathcal{W}(y)&=&-\frac{2 y^2
\Lambda _i}{(d-2) (d-1)}-2 m y^{3-d}+Q^2 y^{-2 (d-3)}+2.
\end{eqnarray}
We evaluate the respective expressions of the second derivative of
effective potential with different matter distributions by using
Eqs.(\ref{27}), (\ref{30}) and (\ref{33}). The corresponding
expressions of the second derivative of the effective potential for
different matter distributions are analyzed graphically in Figures
\textbf{7-9}. The thin-shell gravastar represents an unstable
structure for every choice of matter distribution with different
values of $d$ while for higher values of $d$, the rate of the
unstable structure decreases. For this choice of exterior geometry,
the developed structure expresses unstable structure only.
\begin{figure}\centering
\epsfig{file=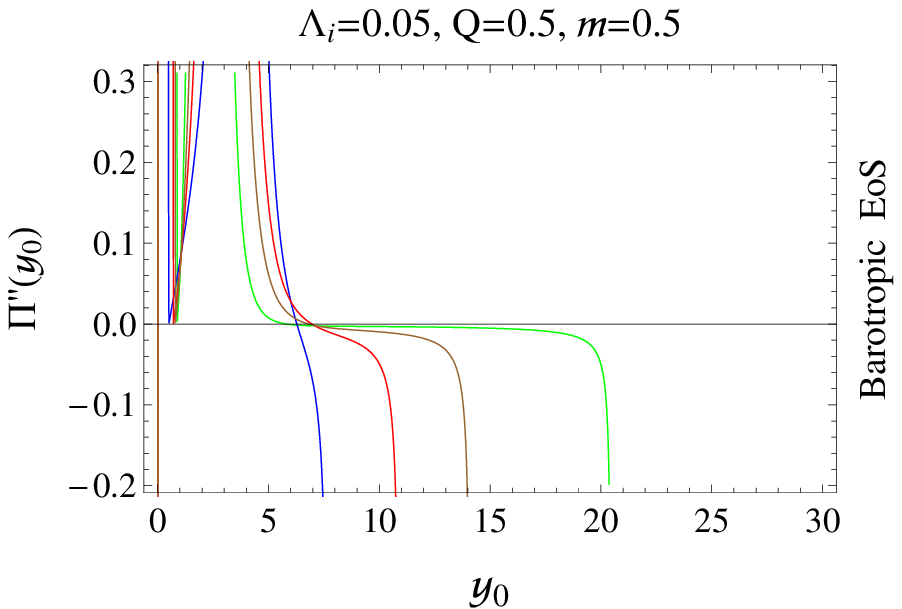,width=.5\linewidth} \caption{Unstable
configuration of $d$-dimensional gravastar for barotropic EoS with
exterior $d$-dimensional RN BH.}
\epsfig{file=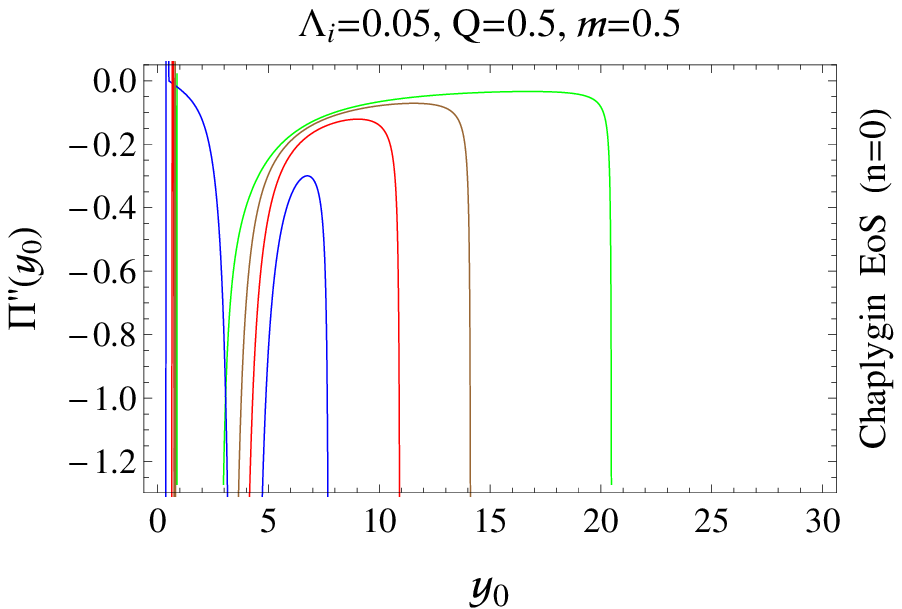,width=.5\linewidth}\epsfig{file=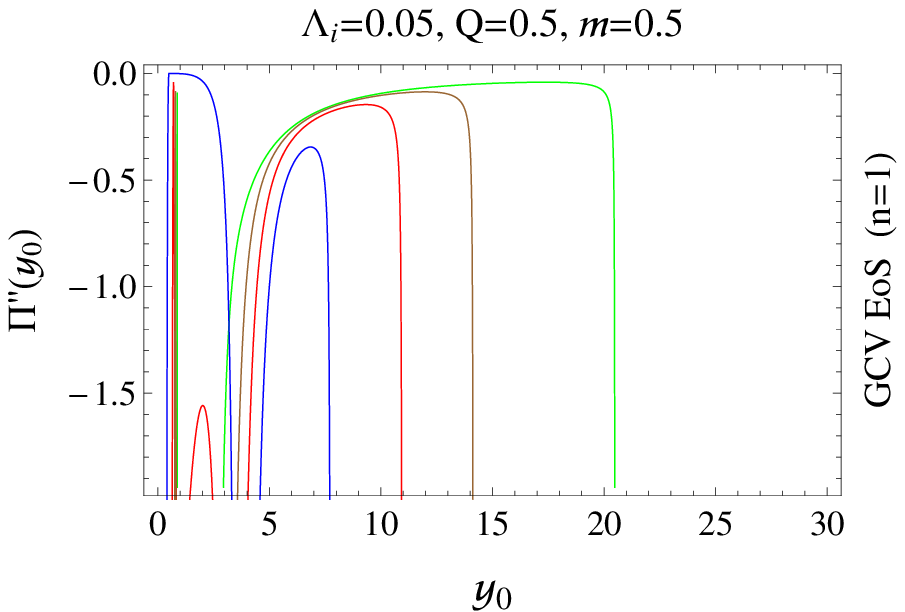,width=.5\linewidth}
\caption{Unstable configuration of $d$-dimensional gravastar for
Chaplygin and GCV EoS with exterior $d$-dimensional RN BH.}
\epsfig{file=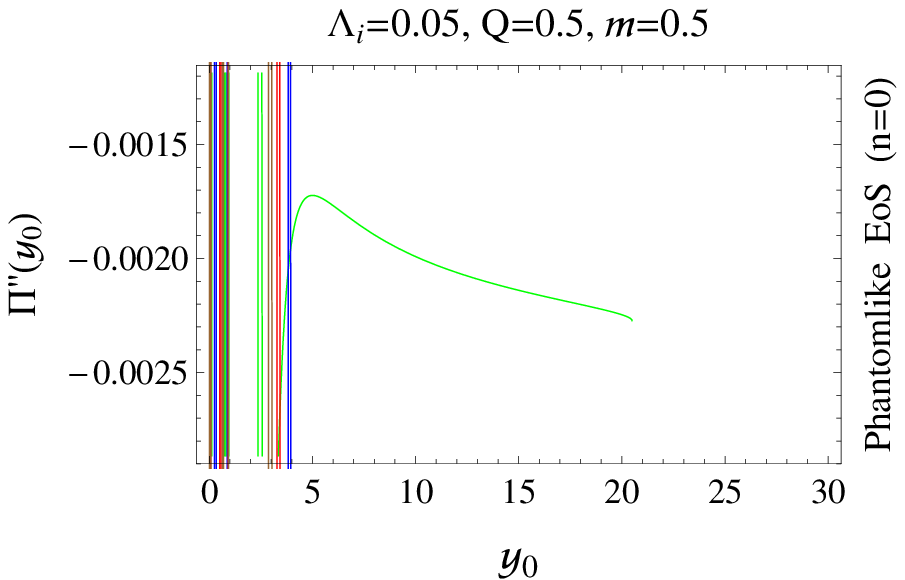,width=.5\linewidth}\epsfig{file=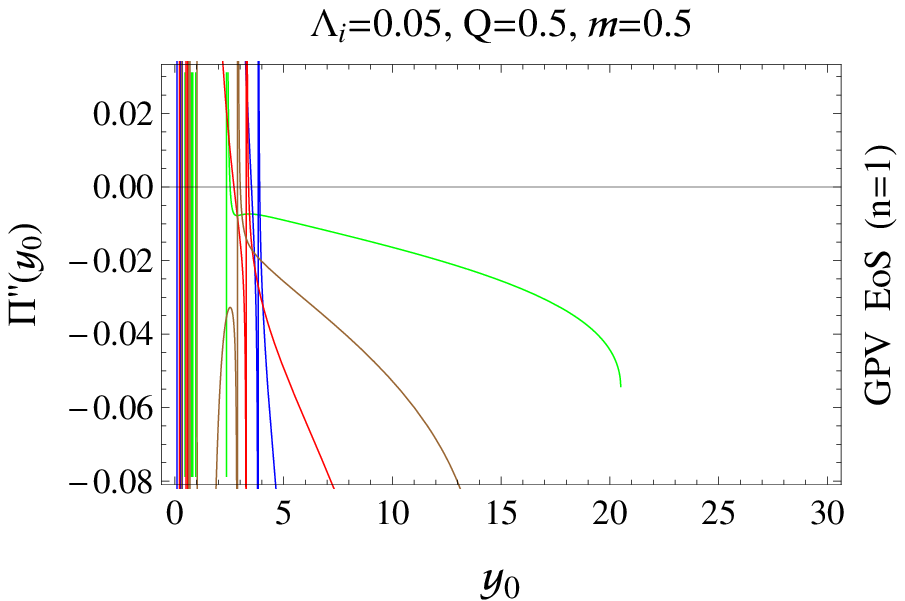,width=.5\linewidth}
\caption{Unstable configuration of $d$-dimensional gravastar for
phantomlike and GPV EoS with exterior $d$-dimensional RN BH.}
\end{figure}

\subsubsection{$d$-dimensional Reissner-Nordstr\"om-dS BH}
\begin{figure}\centering
\epsfig{file=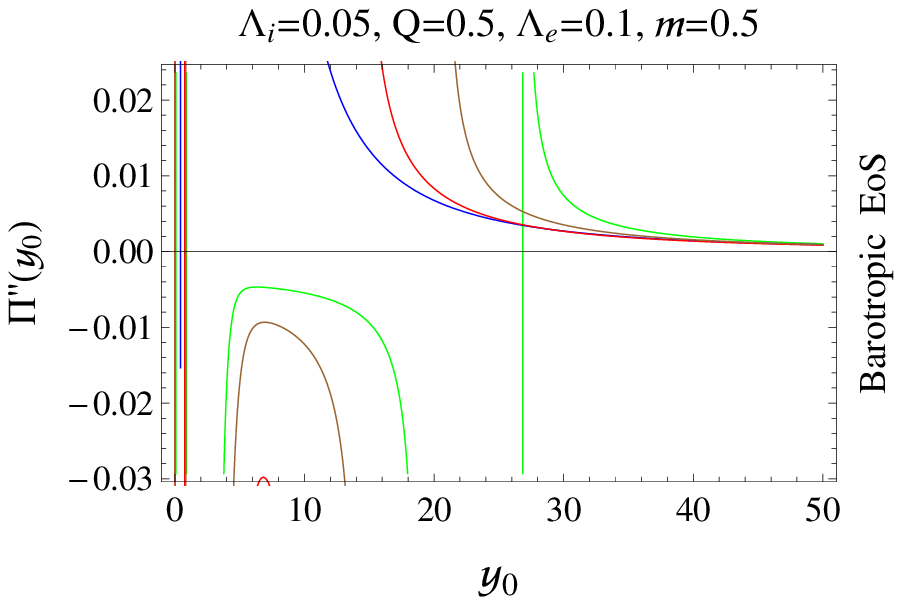,width=.5\linewidth} \caption{Stability of
$d$-dimensional gravastar for barotropic EoS with exterior
$d$-dimensional RN-dS BH. It is found that developed structure shows
initially unstable then stable behavior for suitable physical
parameters.}
\epsfig{file=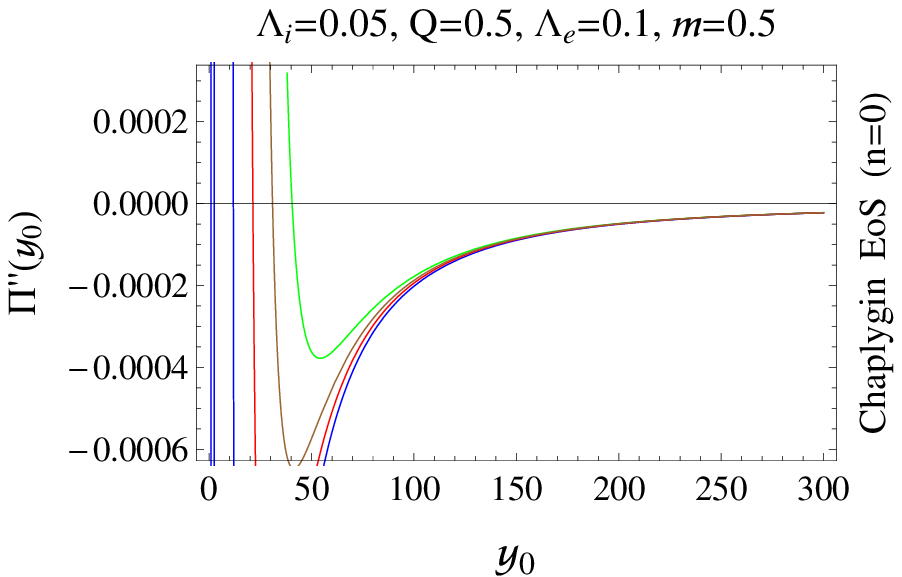,width=.5\linewidth}\epsfig{file=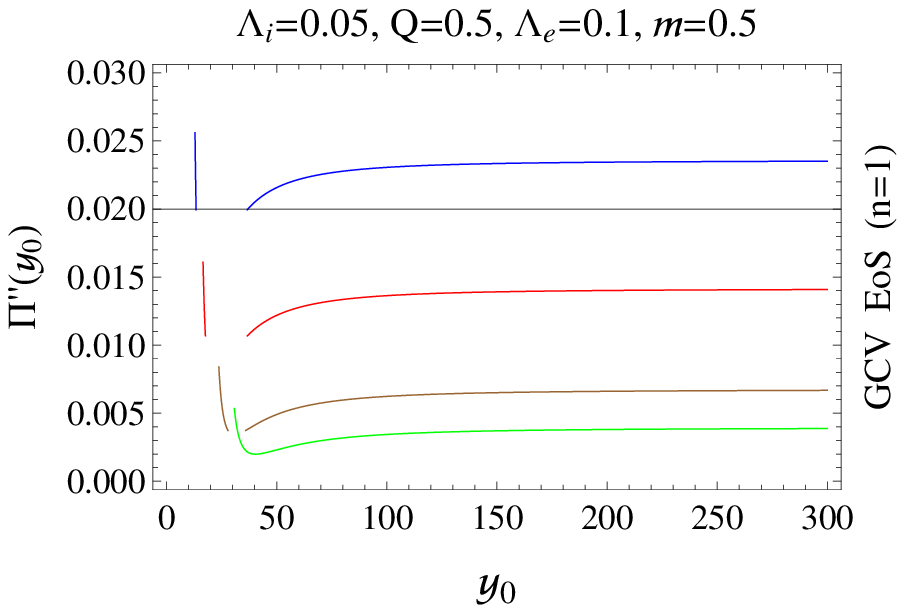,width=.5\linewidth}
\caption{Stability of $d$-dimensional gravastar for Chaplygin and
GCV EoS with exterior $d$-dimensional RN-dS BH. Left plot indicates
unstable behavior for Chaplygin model and right plot shows stable
configuration for GCV EoS.}
\epsfig{file=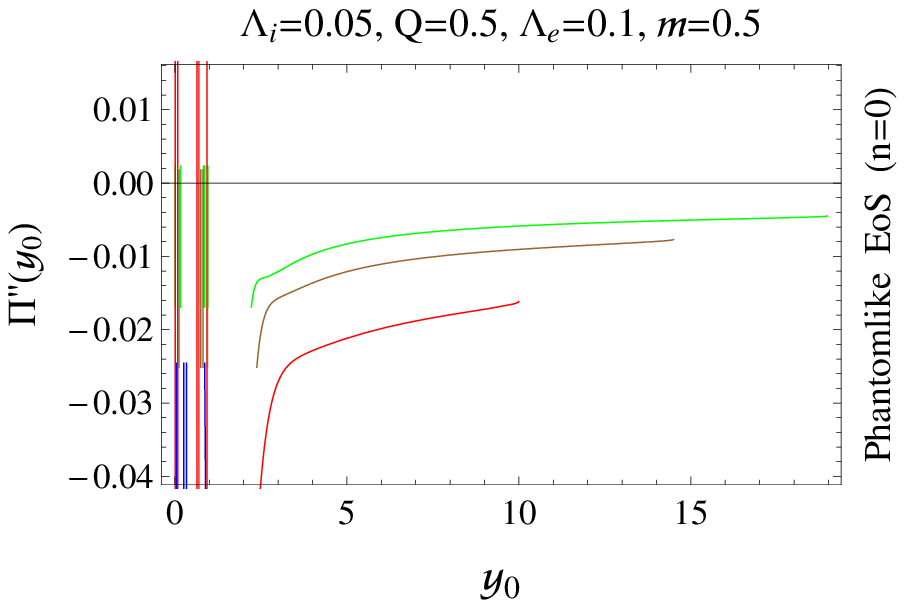,width=.5\linewidth}\epsfig{file=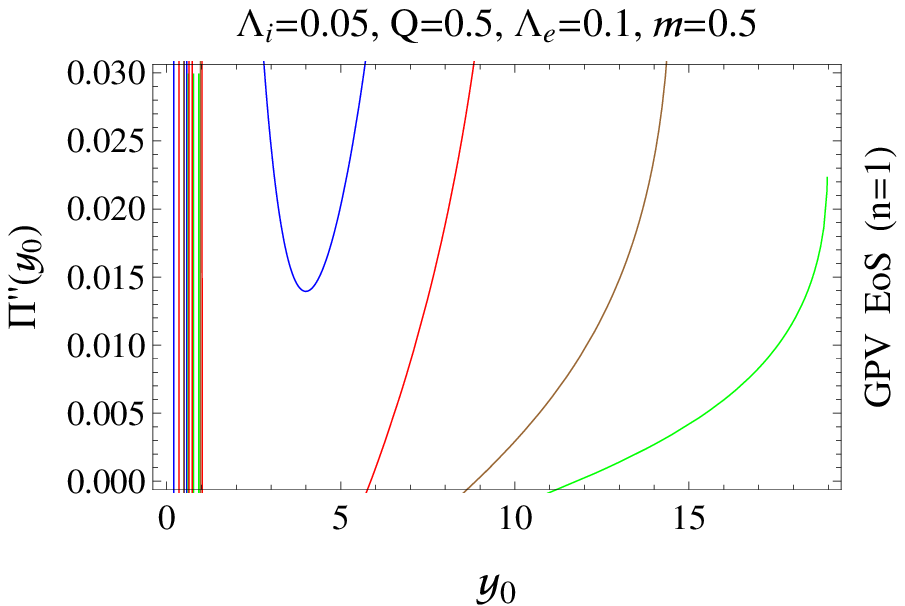,width=.5\linewidth}
\caption{Stability of $d$-dimensional gravastar for phantomlike and
GPV EoS with exterior $d$-dimensional RN-dS BH. Left plot expresses
unstable behavior for phantomlike EoS and right plot represents the
stable geometry for GPV EoS.}
\end{figure}

The respective expressions of $\mathcal{Z}(y)$ and $\mathcal{W}(y)$
are given as
\begin{eqnarray}\nonumber
\mathcal{Z}(y)&=&\frac{2 y^2 \Lambda _e}{(d-2) (d-1)}-\frac{2 y^2
\Lambda _i}{(d-2) (d-1)}+2 m y^{3-d}-Q^2 y^{-2 (d-3)},\\\nonumber
\mathcal{W}(y)&=&2-\frac{2 y^2 \Lambda _e}{(d-2) (d-1)}-\frac{2 y^2
\Lambda _i}{(d-2) (d-1)}-2 m y^{3-d}+Q^2 y^{-2 (d-3)}.
\end{eqnarray}
The stable configurations are analyzed by using Eqs.(\ref{27}),
(\ref{30}) and (\ref{33}) as shown in Figures \textbf{10-12}. We see
that the developed structure indicates partially stable behavior for
barotropic type fluid distribution (Figure \textbf{10}) and unstable
structure for Chaplygin as well as phantomlike EoS (left plots of
Figures \textbf{11} and \textbf{12}). For GCV and GPV EoS,
thin-shell gravastars show stable configuration for every value of
$d$ (right plots of Figures \textbf{11} and \textbf{12}).

\section{Conclusions}

This paper studies stable configurations of gravastars through
linearized radial perturbations for higher dimensions. We consider
higher-dimensional BH spacetimes in the exterior of gravastars and
explore their stability for three different types of matter
distributions, i.e., barotropic, GCV, and GPV EoS. We have developed
a geometrical structure of $d$-dimensional gravastars through the
joining of inner $d$-dimensional dS and outer $d$-dimensional
spherically symmetric BH spacetimes by using the cut and paste
approach. We have considered Lanczos equations to obtain matter
contents located at the hypersurface. The linearized radial
perturbation is applied to the potential function of thin-shell to
explore stable configurations.

Firstly, we have studied stable configurations of gravastar
structure in the background of exterior $d$-dimensional
Schwarzschild BH. We have found that the developed structure
expresses unstable configurations for every type of matter
distribution (Figures \textbf{1-3}). For the exterior
$d$-dimensional Schwarzschild-dS BH, we have obtained unstable
structure for barotropic, Chaplygin, and phantomlike models while
stable structure is found for GCV and GPV models for every suitable
value of physical parameters (Figures \textbf{4-6}).

For the $d$-dimensional RN manifold, the developed framework
represents unstable configurations (Figures \textbf{7-9}) while for
$d$-dimensional RN-dS manifold, the structure indicates unstable
behavior (for barotropic, Chaplygin, and phantomlike models) but
stable configurations for GCV and GPV models (Figures
\textbf{10-12}). Hence, the $d$-dimensional Schwarzschild/RN-dS
geometries lead to stable structure as compared to $d$-dimensional
Schwarzschild/RN BHs. We conclude that the presence of cosmological
constant in the exterior of gravastar geometry leads to stable
structure for GCV and GPV type fluid distributions. It is worthwhile
to mention here that stable configuration of higher-dimensional
gravastar has similar behavior to four-dimensional gravastar.


\begin{thebibliography}{43}

\bibitem{1} P. Mazur, E. Mottola, Gravitational condensate stars: An alternative to black holes, arXiv:
gr-qc/0109035.

\bibitem{2} P. Mazur, E. Mottola, Gravitational vacuum condensate stars, Proc. Nat. Acad. Sci.
101 (2004) 9545.

\bibitem{2a}  M. Visser, S. Kar, N. Dadhich, Traversable wormholes
with arbitrarily small energy condition violations, Phys. Rev. Lett.
90 (2003) 201102.

\bibitem{2aa} M. Sharif, F. Javed, Collapse and expansion of scalar thin-shell
for a class of black holes, Int. J. Mod. Phys. D 28 (2019) 1950046;
Dynamical evolution of scalar field thin-shell for rotating regular
black holes, Ann. Phys. 407 (2019) 198; Stability of
Einstein-power-Maxwell (2+1)-dimensional wormholes, Chin. J. Phys.
61 (2019) 262; Dynamics of scalar shell for rotating and charged
rotating BTZ black holes, Mod. Phys. Lett. A 35 (2019) 1950350.

\bibitem{2aaa} M. Sharif, F. Javed, Stability of charged
Kiselev thin-shell wormholes, Int. J. Mod. Phys. A 35 (2020)
2040015; Stability of charged rotating (2 + 1)-dimensional
wormholes, Int. J. Mod. Phys. D 29 (2020) 2050007; Mechanical
stability of a class of regular thin-shell wormholes, Mod. Phys.
Lett. A 39 (2020) 2050309; Dynamics of thin-shell wormholes with
rotational effects, Int. J. Mod. Phys. A 35 (2020) 2050030.

\bibitem{3} M. Visser, D.L. Wiltshire, Stable gravastars an alternative to black holes?, Class. Quantum Grav.
21 (2004) 1135.

\bibitem{3b} D. Horvat, S. Ilijic, Gravastar energy conditions revisited, Class. Quantum Grav.
24 (2007) 5637.

\bibitem{4} B.M.N. Carter, Stable gravastars with generalized exteriors, Class. Quantum Grav.
22 (2005) 4551.

\bibitem{5} D. Horvat, S. Sasa Ilijic, A. Marunovic, Electrically charged gravastar
configurations, Class. Quantum Grav. 26 (2009) 025003.

\bibitem{6} F. Rahaman, A.A. Usmani, S. Ray, S. Islam, The (2+1)-dimensional gravastars, Phys. Lett.
B 707 (2012) 319;  The (2+1)-dimensional charged gravastars, ibid.
717 (2012) 1.

\bibitem{6a} C.F.C. Brandt, et al., Charged gravastar in a dark energy universe, J. Mod. Phys. 4 (2013) 869.

\bibitem{8} P. Rocha, et al., Bounded excursion stable gravastars
and black holes, J. Cosmol. Astropart. Phys. 06 (2008) 25; R. Chan,
M.F.A. da Silva, P. Rocha, A. Wang, Stable gravastars of anisotropic
dark energy, J. Cosmol. Astropart. Phys. 03 (2009) 10.

\bibitem{14} R. Chan, M.F.A. da Silva, P. Rocha, How the cosmological constant affects
gravastar formation, J. Cosmol. Astropart. Phys. 12 (2009) 17.

\bibitem{15} R. Chan, M.F.A. da Silva, How the charge can affect the
formation of gravastars, J. Cosmol. Astropart. Phys. 07 (2010) 29.

\bibitem{15cc} F.S.N. Lobo, A.V.B. Arellano, Gravastars supported by nonlinear electrodynamics, Class. Quantum Grav.
24 (2007) 1069.

\bibitem{10} F.S.N. Lobo, R. Garattini, Linearized stability analysis
of gravastars in noncommutative geometry, J. High Energy Phys. 1312
(2013) 065.

\bibitem{11} A. \"{O}vg\"{u}n, A. Banerjee, K. Jusufi, Charged thin-shell
gravastars in noncommutative geometry, Eur. Phys. J. C 77 (2017)
566.

\bibitem{12} M. Sharif, F. Javed, Stability of gravastars with exterior regular black holes, Ann. Phys.
415 (2020) 168124; Stability of charged thin-shell gravastars with
quintessence, Eur. Phys. J. C 81 (2021) 47.

\bibitem{14a} A.A. Usmani, et al., Charged gravastars admitting conformal motion, Phys. Lett. B
701 (2011) 388.

\bibitem{14b} P. Bhar, Higher dimensional charged gravastar admitting
conformal motion, Astrophys. Space Sci. 354 (2014) 457.

\bibitem{14c} Ghosh, S., Rahaman, F., Guha, B.K. and Ray, S., Charged
gravastars in higher dimensions, Phys. Lett. B 767 (2017) 380;
Gravastars with higher dimensional spacetimes, Ann. Phys. 394 (2018)
230.

\bibitem{13d} G.A.S. Dias, J.P.S. Lemos, Thin-shell wormholes in
$d$-dimensional general relativity: Solutions, properties, and
stability, Phys. Rev. D 82 (2010) 084023.

\bibitem{13g} E.F. Eiroa, C. Simeone, Aspects of spherical
shells in a $d$-dimensional background, Int. J. Mod. Phys. D 21
(2012) 1250033.

\bibitem{13gg} A. Banerjee, K. Jusufi, S. Bahamonde, Stability of a
$d$-dimensional thin-shell wormhole surrounded by quintessence,
Gravit. Cosmol. 24 (2018) 1.

\bibitem{13j} M. Sharif, F. Javed, Dynamics of the scalar shell in higher dimensions, Ann. Phys.
416 (2020) 168146; Stability and dynamics of regular thin-shell
gravastars, J. Exp. Theor. Phys. 132 (2021) 381.

\bibitem{12f} R.A. Konoplya, A. Zhidenko, Stability of multidimensional black
holes: Complete numerical analysis, Nucl. Phys. B 777 (2007) 182;
Stability of higher dimensional Reissner-Nordstr\"om-anti-de Sitter
black holes, Phys. Rev. D 78 (2008) 104017.

\bibitem{20} V. Varela, Note on linearized stability of Schwarzschild
thin-shell wormholes with variable equations of state, Phys. Rev. D
92 (2015) 044002.

\bibitem{20a} M. Sharif, F. Javed, On the stability of bardeen thin-shell wormholes, Gen. Relativ. Gravit.
48 (2016) 158; Linearized stability of Bardeen anti-de Sitter
wormholes, Astrophys. Space. Sci. 364 (2019) 179. Stability of
charged thin-shell and thin-shell wormholes: A comparison, Phys.
Scr. 96 (2021) 055003.

\bibitem{21} E.F. Eiroa, C. Simeone, Stability of Chaplygin gas thin-shell wormholes, Phys. Rev. D
76 (2007) 024021.

\bibitem{22} P.K.F. Kuhfittig, The stability of thin-shell wormholes with a
phantom-like equation of state, Acta Phys. Polon. B 41 (2010) 2017.

\end{thebibliography}
\end{document}